# Sputtered Gold Nanoparticles Enhanced Quantum Dot Light-emitting Diodes[*]

Abida Perveen[1], Xin Zhang(张欣)[2], Jialun Tang(汤加仑)[2], Dengbao Han(韩登宝)[2], Shuai Chang(常帅)[2,†], Luogen Deng(邓罗根)[1], Wenyu Ji(纪文宇)[3], Haizheng Zhong(钟海政)[2,†]

[1] *Department of Physics, Beijing Institute of Technology, Beijing, 100081, China*

[2] *Department of Materials Science and Engineering, Beijing Institute of Technology, Beijing, 100081, China*

[3] *Department of Physics, Jilin University, Changchun, 130023, China*

Surface plasmonic effects of metallic particles have been known to be an effective method to improve the performance light emitting didoes. In this work, we reported the sputtered Au nanoparticles enhanced electroluminescence in inverted quantum dot light emitting diodes (ITO/Au NPs/ZnMgO/QDs/TFB/PEDOT:PSS/Al). By combining the time-resolved photoluminescence, transient electroluminescence and ultraviolet photoelectron spectrometer measurements, the enhancement can be explained to the internal field enhanced exciton coupling to surface plasmons and the increased electron injection rate with Au nanoparticles incorporation. Phenomenological numerical calculations indicated that the electron mobility of the electron transport layer was increased from $1.39 \times 10^{-5}$ to $1.91 \times 10^{-5}$ cm$^2$/V·s for Au NPs modified devices. As a result, the maximum device luminescence is enhanced by 1.41 folds from 14,600 to 20,720 cd/cm$^2$ and maximum current efficiency is improved by 1.29 folds from 3.12 to 4.02 cd/A.

**Keywords: gold nanoparticles, plasmonic effect, quantum dots, light-emitting diodes**

**PACS:** 61.46.Df, 73.20.Mf

## 1. Introduction

Quantum dots (QDs) present attractive features of precise emission bandwidth, saturated emission, tunable emission wavelengths, high quality production with low cost solution processing.[1-5] Such promising features make them potential candidates for next generation display technologies through fully functionalized QD-LEDs.[6-10] Since the demonstration of first QD-LED in 1994, continuous efforts have been made to improve their performance and the device is comparable with conventional organic light-emitting diodes in a number of performance factors.[2-3,11] In particular, the inverted device is suitable for the display due to the compatibility to the n-channel TFT. QD-LEDs still suffer from low luminous efficiency and poor stability [12,13] due to imbalanced transportation and injection of carriers[14], intrinsic photoluminescence quenching[15], and weak out-coupling.[16,17]

[*] Project supported by the National Natural Science Foundation of China (21603012, 61735004, 61722502)

[†] Corresponding author. E-mail: schang@bit.edu.cn

[†] Corresponding author. E-mail: hzzhong@bit.edu.cn

Surface plasmons (SPs), which are collectively oscillating free electrons at the interface of metal and dielectric,[8,18] are being extensively investigated[10,14,19,20] and being preliminarily investigated in QD-LEDs.[21] Enhancements such as PL[8,22] and internal quantum efficiency (IQE)[19] of electroluminescent devices have been demonstrated. Furthermore, localized surface plasmons (LSPs) associated with noble metal nanostructures show resonance coupling to excitons providing enlargement of local electromagnetic field, which results in rapid radiative emission rates by effective energy coupling from QDs to LSPs.[8,23] Metal nanostructures have been previously made[3] and incorporated by chemical or thermal vapor deposition of thin films with post annealing at high temperature[1], by spin coating[24], or by patterned tempelates.[16] However most methods to incorporate plasmonic metal nanostructures require complicated process and sometimes incompatible with solution processed device applications.

Herein, directly sputtered gold nanoparticles (Au NPs) on ITO substrate is proposed for the enhancement of QD-LEDs. The generated Au NPs layer creates plasmonic effect without damaging soft and thin underlying organic films[25] and contributes to increase not only the work function of electron transport layer, but also the quick electron injection into QDs and rapid coupling of excitons to LSPs.

## 2. Experimental Section
### 2.1 Preparation of Au NP film
Au target film was set 5 cm above the aimed substrate. The chamber was vacuumized to 10 Pa and the sputtering process was initiated at 10 mA with various time span of 5 sec, 10 secs, 20 secs and 30 secs, respectively. Then the sputtered Au films were annealed at 280 °C for 20 min to form discrete NPs.

### 2.2 Device fabrication
Mg doped ZnO (ZnMgO) was spun coated on the ITO substrate at 2000 rpm for 60 secs and annealed at 110 °C for 20 min. CdSe QDs were dispersed in n-heptane with a concentration of 30 mg/ml and spun coated at 2500 rpm for 60 secs, followed by heating to 90 °C for 10 min. Hole transport layer poly(9,9-dioctylfluorene-co-N-(4-(3-methylpropyl))diphenylamine) (TFB) was prepared in chlorobenzene and spun coating at 4000 rpm for 30 secs and annealed at 150 °C for 30 min, followed by spinning coating of PEDOT:PSS at 4000rpm for 60 secs and baked at 160 °C for 15 min. Finally, 100 nm Al was thermally evaporated as electrode.

## 2.3 Characterizations

Scanning electron microscopy (SEM) images and energy dispersive spectra (EDS) were taken on a S-4800 microscopy (Hitachi, Ltd., Japan). Ultraviolet–visible spectroscopy (UV-Vis) absorption spectra of QDs solutions were measured on a UV-6100 spectrophotometer and PL spectra were obtained using an F-380 fluorescence spectrometer. Current density-voltage (J-V) characteristics of QD-LEDs were measured using a Keithley 2400 power source analyzer. The luminance was measured using a spectroradiometer (Photo Research Inc. PR-655). Time-resolved photoluminescence (TRPL) measurement was collected using fluorescence lifetime measurement system (Edinburgh FL920) at an excitation wavelength of 405 nm. Time-resolved electroluminescence (TREL) was carried out by applying a pulsed voltage to the device, then the output signal was collected by an avalanche photo diode (APD) detector (Hamamatsu C10508-01) and exported on a digital oscilloscope (Tektronix DPO7104C). Ultraviolet photoelectron spectrometer (UPS) was carried out on a ESCALAB 250XI XPS Microprobe (with a UV source accessory, ThermoFisher Scientific) using He I$\alpha$ photon energy (21.22eV).

## 3. Results and Discussion

The corresponding SEM micrograph of optimized sputtered substrate after thermal annealing is presented in Figure 1 (a) and Figure S1. It can be clearly seen that, spherical Au NPs with average size of 22 nm (20 second sputtering time) are uniformly distributed over the entire surface of ITO. The change in the size of as-prepared Au NPs were found to be 12 nm, 13.5 nm, 22 nm and 35 nm corresponds to the alteration of sputtering time. Figure 1(b) shows the size dependent extinction characteristics of Au NPs on ITO substrates and the PL spectra of QD film. Size dependent absorption of Au NP films was observed and the absorption of non-annealed continuous sputtered metal films showed no plasmonic resonance absorption peak. PL emission peak of QDs at the wavelength of 518 nm matches perfectly with absorption peak of 22 nm Au NP films, which indicates highly efficient coupling of excitons to the SPs.[9]

To incorporate the Au plasmonic layer into the electroluminescent QD-LED device, all-solution processed inverted device structure of ITO/Au NPs/ZnMgO/QDs/TFB/PEDOT:PSS/Al was applied as shown in Figure 2(a). Where ZnMgO has been used as electron transport layer and also plays the role of dielectric spacer layer between Au NP and QDs, which diminishes non-radiative quenching induced by SPs.[26-29] Figure 2(b) depicts energy level diagram, according to which PEDOT:PSS and TFB are performing roles of hole injecting and hole transport layers, respectively. Figure 2(c) elaborates PL spectra of Au/ZnMgO/QDs and ZnMgO/QDs films, 1.7 folds of PL enhancement was observed for plasmonic enhanced QD films as compared to that of pure QD films, signifying the enhanced spontaneous emission through SP coupling. In order to certify the origin of PL intensity enhancement in Au NPs existence, we performed TRPL measurements for QD films coated over pure ZnMgO and Au NP/ZnMgO substrates, corresponding decay for both films are shown in Figure 2(d). It has been revealed that excitons are coupling to SPs more rapidly than spontaneous recombination of excitons.[30,31] By the incorporation of Au NPs the lifetime of PL should be reduced. Calculations depict the decay times to be 24.82 ns and 23.38 ns for normal and plasmonic QD films, respectively. Reduction in the decay time is related to the increase in exciton coupling, and resulting in enhanced QD emission.[32] This new way of recombination in which electron-hole pairs (excitons) couple to vibrating electrons on the surface of metal to produce SPs instead of phonons or photons accelerates spontaneous radiative emission rates and IQEs.[19,33]

Half normal/half plasmonic device was fabricated on the same substrate to diminish the systematic error among different batches of devices, as shown in Figure 3(a). A comparison image of normal and plasmonic devices driven at 5 V was also presented. Figure (S2) shows normalized EL and PL emission at the same peak wavelength of 518 nm. Luminance-voltage, current density-voltage characteristics of the normal and plasmonic devices are displayed in Figure 3(b). The plasmonic device displays highest luminance of 20,720 cd/cm$^2$, 1.41 folds larger as compared to normal device. Maximum current efficiency of 4.02 cd/A for plasmonic device is obtained (Figure 3(c)), which is improved by 1.29 folds from normal device. The improvement of current efficiency can be ascribed to the internal field enhancement in the presence of Au NPs due to resonance coupling and fast injection of charge carriers, which ultimately leads to more balanced charge injection. As a result, the turn on voltage plasmonic device is reduced from 3 V to 2.8 V.

TREL was performed to elaborate transportation of charge carriers under the effect of Au NPs (Figure S3). Transient evolution profiles of normalized electroluminescence (EL) were obtained under a 5 µs pulse with a pulse cycle of 1ms. Due to low power of per pulse, a relatively high applied voltage of 7.5 V was applied to yield a measurable luminance of the device. It is observed that EL evolution profiles of normal and plasmonic devices experienced a delay period first, followed by a fast rise due to enhanced spontaneous emission and finally overtaken by a slow rise until it reaches to its maximum value. It is clearly seen from the comparison of EL evolution profiles of plasmonic and normal devices that normal device takes longer delay and shows slower initial rise and experiences longer time lag in reaching to maximum EL than that of plasmonic device.[33] A mobility determination technique was applied which is based on phenomenological numerical model.[33]

From plots of normalized {[EL(t→∞)−EL(t)/EL(t→∞)]} vs time and EL vs time (Figure 4 (a)), delay time td is defined as the intersection of the horizontal regime and the first slope of the curve, initial rise time t1 is the time where the first and second slopes of the curve intersects and t2 is the time when the EL intensity reaches 95% of its steady value.

We assume that $\mu_h \approx \mu_e$, and taking time measurements $t_d$, $t_1$ and $t_2$ from our experimental measurements we take equation $\mu_e = L^2/(V-V_{bi})(t_d+t_1)$ to calculate mobility of electrons[33], where $V_{bi}$ is assumed built in voltage of devices, taken from the difference between work functions of ITO to Au NPs and ZnMgO NPs (1.04 V and 0.93 V for normal and plasmonic devices, respectively) and L is thickness of the device (120 nm).

TABLE I: Characteristics times $t_d$, $t_1$, $t_2$ and electron mobility µe

| Normalized EL | $t_d$ (µs) | $t_1$ (µs) | $t_2$ (µs) | $\mu_e$ (cm$^2$/V·s) |
|---|---|---|---|---|
| W/O Au | 0.77 | 0.84 | 2.05 | $1.39e^{-5}$ |
| With Au | 0.50 | 0.65 | 1.65 | $1.91e^{-5}$ |

UPS spectrum of ITO/ZnMgO and ITO/Au/ZnMgO is shown in Figure 4 (b). All the spectrum was measured using He Iα photons (hυ=21.22 eV). The LOMO energy levels calculated from UPS characteristics (Table S1) show a 0.11 eV decrement for the ITO/Au/ZnMgO film as compared with the ITO/ZnMgO film. Therefore, the charge injection barrier between the ZnMgO and QDs is reduced with Au incorporation and such reduction consequently contributes to the balanced transportation of carriers across the QDs layer. As a result, delay time td is shortened by 0.27 µs for plasmonic device compared to normal device, manifesting faster injection of charge carriers from the charge transfer layers to QDs.

Time for initial rise, $t_1$ is shortened by 0.19 µs for plasmonic device. This can be attributes to the efficient resonance coupling of excitons to SPs, which in turn speed up spontaneous emission with continuous balance flow of carriers in this region. Therefore, the emission process is accelerated by resonance coupling to produce brighter emission peak as well. The comparison of $t_2$ reveals that plasmonic device reaches to 95% of its steady EL intensity 0.4 µs faster than that of normal device. This consistent behavior of EL evolution stands by our assumption of $\mu_h \approx \mu_e$ across the device. Due to this balance in transportation of charge carriers, mobility enhancement from 1.39x10$^{-5}$ to 1.91x10$^{-5}$ cm$^2$/V·s for Au NPs containing devices has been achieved. The I-V traces of the devices followed the Mott-Gurney's power law (Figure S4) with an Ohmic region at the lower bias and a space charge-limited current (SCLC) regime at higher bias. For plasmonic devices, shorter Ohmic region was observed compared to those of normal devices, which signifies more rapid carrier transportation, efficient resonance coupling enhanced spontaneous emission and reduced quenching of excitons under the influence of plasmonic effect. Furthermore, electron only devices with the structure of ITO/ZnMgO/QDs/Al and ITO/Au NPs/ZnMgO/QDs/Al were fabricated and the corresponding current density of the latter with Au NPs is higher than that of the former (Figure S5), demonstrating improved charge injection rate from charge transport layer to QDs. Combined with the TREL results, the reduced time delay and efficient transportation results in balance and smooth transportation of carriers in plasmonic devices.

4. **Conclusion and perspectives**

In summary, we elaborated SP enhanced QD-LEDs using Au NPs. Enhancements in the device performance are attributed to the increased charge injection rates into QDs, balanced transportation of carriers across the device and the strong resonance coupling of SPs and excitons verified by TRPL, TREL, UPS and phenomenological numerical results. Enhancement in carrier mobility from $1.39 \times 10^{-5}$ to $1.91 \times 10^{-5}$ $cm^2/V \cdot s$ verifies the improvement in transportation of charge carriers. Based on the Au incorporation, efficient green emission QD-LEDs with a maximum luminance of 20,700 $cd/m^2$ and a maximum current efficiency of 4.02 cd/A were achieved. These results demonstrate an effective method to apply plasmonic metal nanostructures for achieving super bright and efficient solid-state light emitting devices.

**Acknowledgment**

The authors want to thank BOE Technology Group Co., Ltd. for providing quantum dot samples.

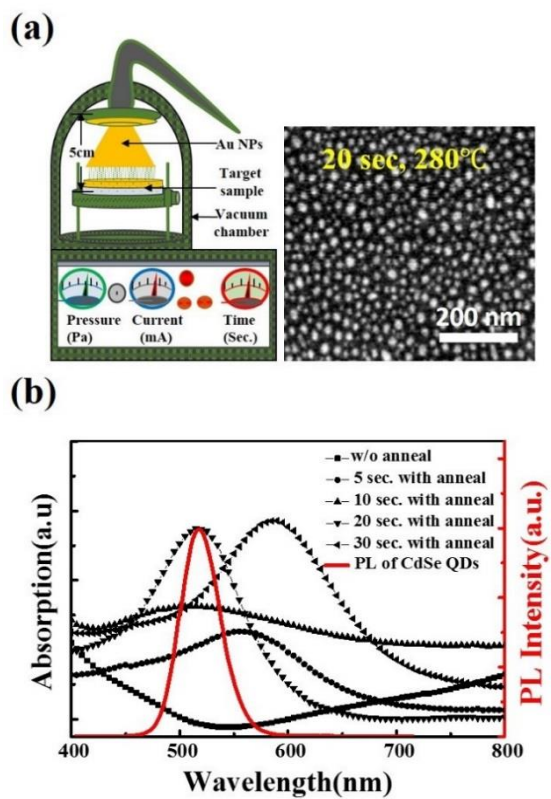

FIG. 1. (a) Schematic of Sputtering machine and SEM micrograph of Au NP film; (b) Absorption of different sized Au NPs, showing size dependent peak variation, and PL spectra of CdSe QDs.

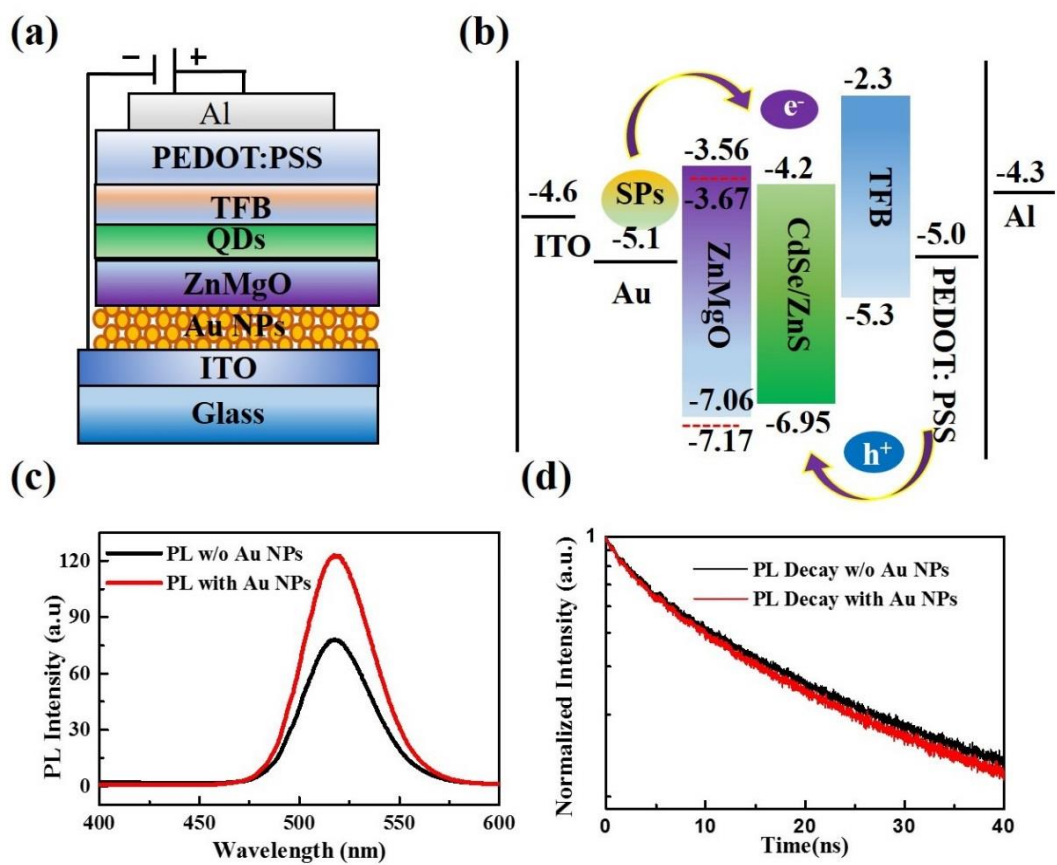

FIG. 2. (a) Illustration of device structure; (b) Energy Level diagram; (c) PL intensities of normal and plasmonic substrates; (d) PL decay behavior of normal and plasmonic substrates.

Chinese Physics B

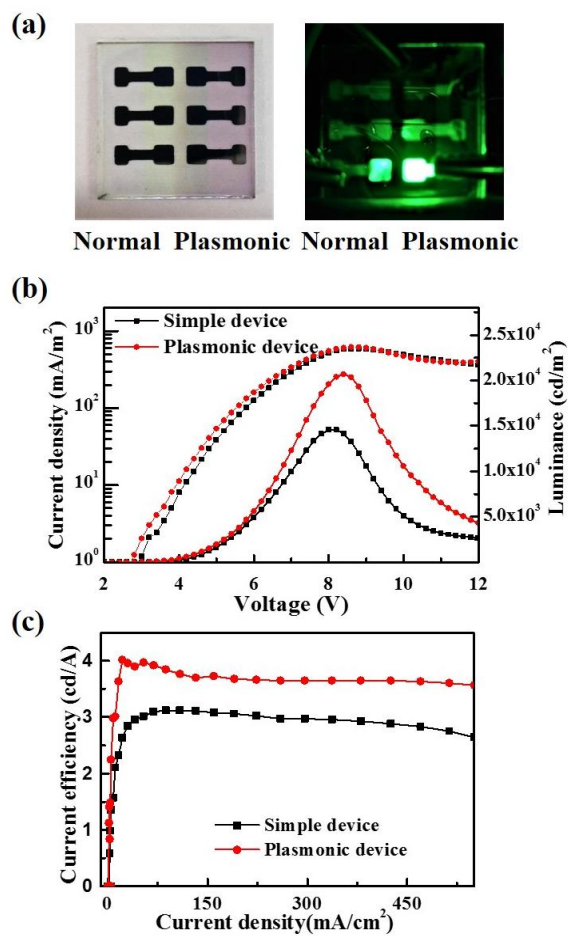

FIG. 3. (a) Optical image of half normal/half plasmonic devices on the same substrate before on after lighting, (b) luminance and current density as a function of applied voltage for simple and plasmonic devices, (c) current efficiency comparison for devices as a function of current density.

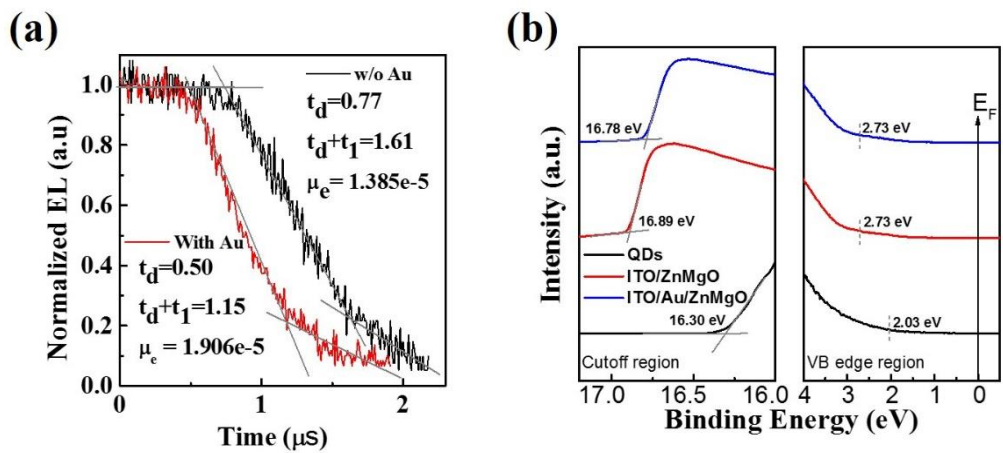

FIG. 4. (a) Plot of simulated normalized EL according to phenomenological model, showing behavior of normalized EL decay profile showing enhanced carriers mobility as a function of time. (b) UPS spectrum of QD, ZnMgO and Au NP enhanced ZnMgO films.

# Supporting Information

**Sputtered Gold Nanoparticles Enhanced Quantum Dot Light-emitting Diodes**

Abida Perveen[1], Xin Zhang(张欣)[2], Jialun Tang(汤加仑)[2], Dengbao Han(韩登宝)[2], Shuai Chang(常帅)[2], Luogen Deng(邓罗根)[1], Wenyu Ji(纪文宇)[3], Haizheng Zhong(钟海政)[2]

[1] *Department of Physics, Beijing Institute of Technology, Beijing, 100081, China*

[2] *Department of Materials Science and Engineering, Beijing Institute of Technology, Beijing, 100081, China*

[3]*Department of Physics, Jilin University, Changchun, 130023, China*

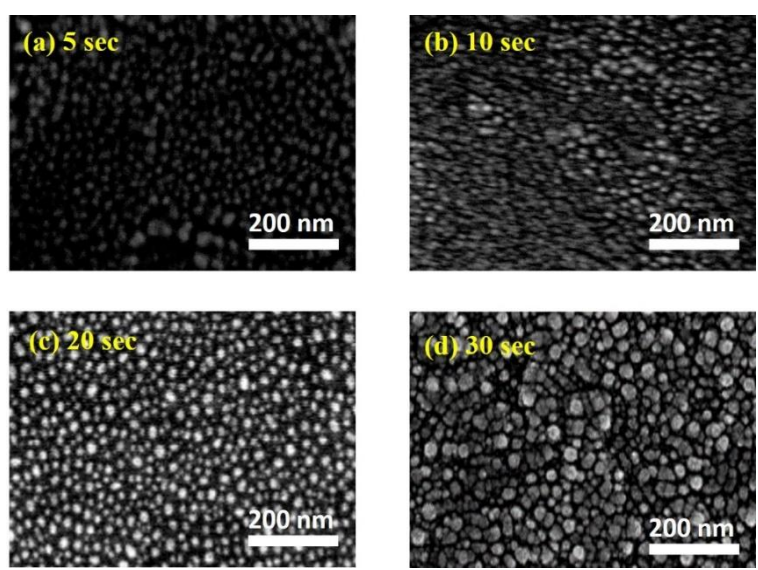

Figure S1: SEM micrographs of sputtered Au from 5 cm height for (a) 5 sec, (b) 10sec, (c) 20sec, (d) 30 sec and annealed at 280°C. Average size of Au NPs for different sputtering times and heights are calculated by Image J software to be 20nm, 20nm, 30nm.

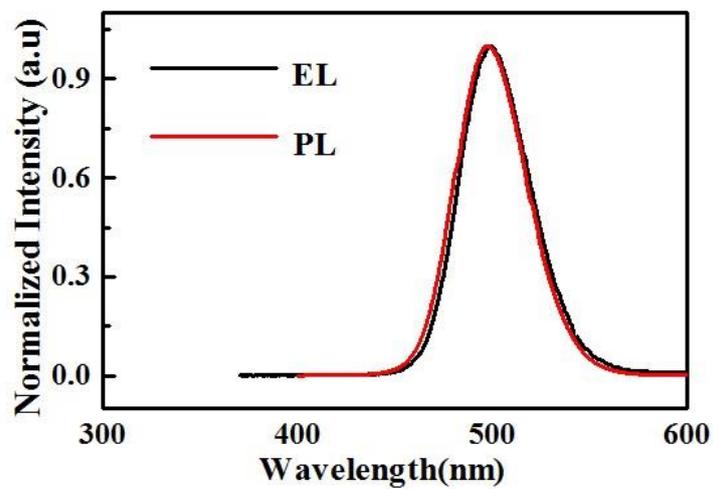

Figure S2: Graph giving comparison of normalized intensities of the devices.

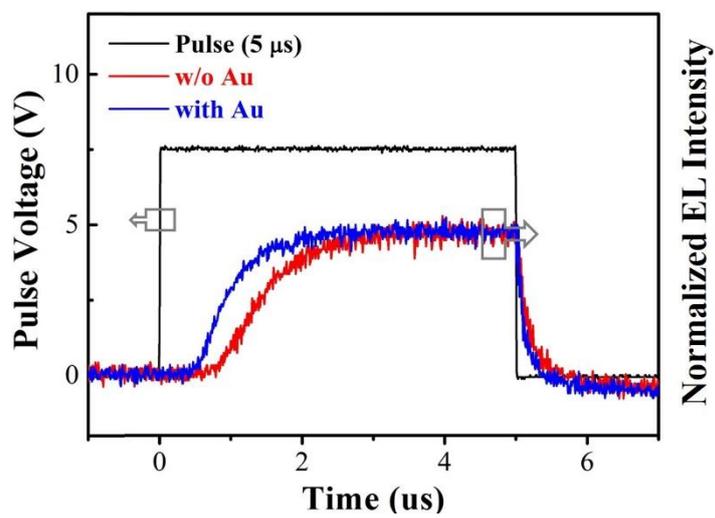

Figure S3: Experimental results for decay of normalized TREL for simple and plasmonic devices under 5 μs pulse voltage.

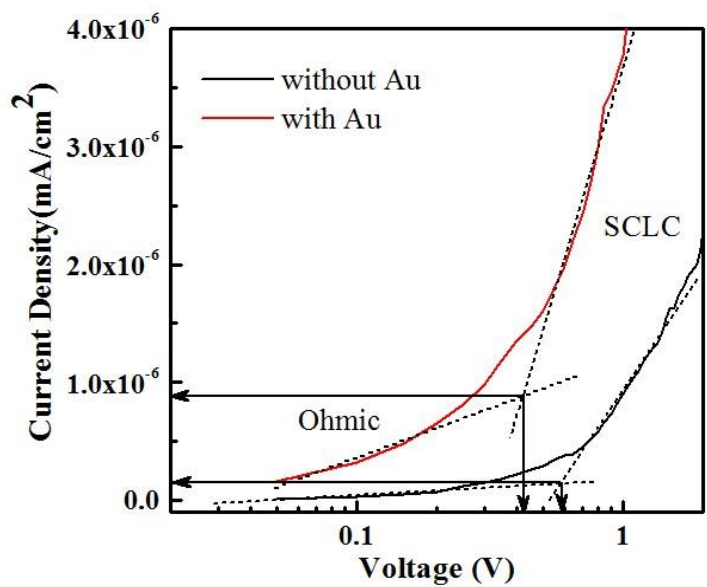

Figure S4: Graph giving current density as a function of applied voltage for simple and plasmonic devices for the voltage range 0.01–2 V.

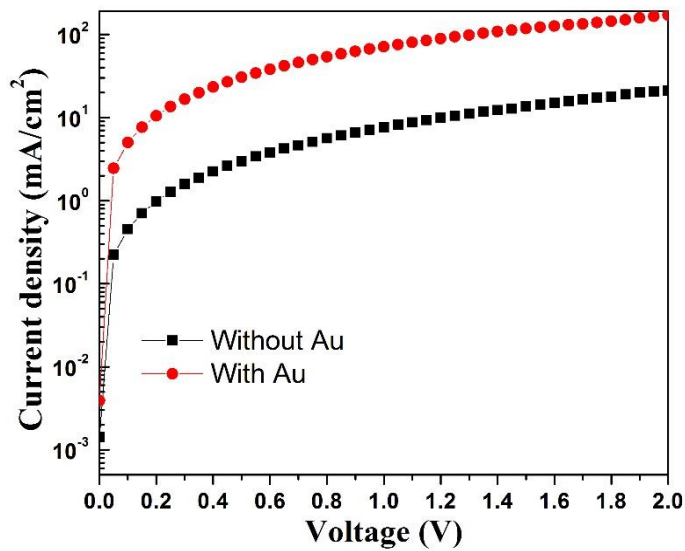

Figure S5: Electron only devices with the structure of ITO/ZnMgO/QDs/Al and ITO/Au NPs/ZnMgO/QDs/Al.

**Table S1:** Work function calculations by UPS measurements.

| Sample | Fermi Energy($E_F$) (eV) | Cut off K.E ($E_{cutoff}$) (eV) | $\phi = h\nu - (E_{cutoff} - E_F)$ $h\nu = 21.22$ (eV) | VB level (eV) | CB level (eV) |
|---|---|---|---|---|---|
| QD | 0 | 16.30 | 4.92 | -6.95 | -4.2 |
| ITO/ZnMgO | 0 | 16.89 | 4.33 | -7.06 | -3.56 |
| ITO/Au/ZnMgO | 0 | 16.78 | 4.44 | -7.17 | -3.67 |